PCCP

COMMUNICATION

View Article Online
View Journal | View Issue

Cite this: *Phys. Chem. Chem. Phys.*, 2016, **18**, 31378

Received 8th September 2016,
Accepted 20th October 2016

DOI: 10.1039/c6cp06222a

www.rsc.org/pccp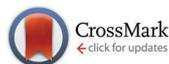

# Hydrogen abstraction from metal surfaces: when electron–hole pair excitations strongly affect hot-atom recombination

Oihana Galparsoro,[abc] Rémi Pétuya,[c] Fabio Busnengo,[d] Joseba Iñaki Juaristi,[cef] Cédric Crespos,[ab] Maite Alducin[cf] and Pascal Larregaray*[ab]Using molecular dynamics simulations, we predict that the inclusion of nonadiabatic electronic excitations influences the dynamics of preadsorbed hydrogen abstraction from the W(110) surface by hydrogen scattering. The hot-atom recombination, which involves hyperthermal diffusion of the impinging atom on the surface, is significantly affected by the dissipation of energy mediated by electron–hole pair excitations at low coverage and low incidence energy. This issue is of importance as this abstraction mechanism is thought to largely contribute to molecular hydrogen formation from metal surfaces.Heterogeneous chemical reactions[1] of hydrogen atoms and molecules are of relevance in many natural and technological contexts, such as interstellar/atmospheric chemistry,[2–5] catalysis,[6–8] semi-conductor processing,[9,10] and plasma-wall interactions.[11,12] Among the involved elementary processes, abstraction of adsorbed species by atom scattering is a key reaction as it affects the density of free surface sites[13] and, depending on the material, may result in exothermic diatom formation, the energy of which can be channeled into highly excited molecules[14–17] or into surface heating.[13]

Numerous theoretical[18–34] and experimental studies[35–44] have concerned hydrogen recombination on metals in the last few decades. When the hydrogen atom chemisorption energy is high, thermal associative desorption *via* the Langmuir–Hinshelwood reaction[45] is ineffective on the picosecond timescale. Rather, hydrogen recombinations may proceed *via* the Eley–Rideal or hot-atom abstraction.[46,47] The former mechanism involves direct collision between a scattering atom (projectile) and an adsorbed species (target), while the latter is governed by hyperthermal diffusion of the projectile onto the surface prior to abstraction. Molecular dynamics simulations[23,24,28] and comparisons between kinetics experiments and models[23,29,39,44] have suggested hydrogen abstraction to essentially proceed *via* the hot-atom process for various metals. Though recent theoretical works on W(110) have confirmed this assertion at low coverage and incidence energies, hot-atom contribution to recombination was predicted to substantially decrease with increasing incidence energy and/or coverage[48] using an electronically adiabatic model. Upon increasing incidence energy, the efficiency for trapping of the impinging hydrogen decreases at the expense of reflection, as the required amount of energy to be withdrawn from the normal motion to the surface increases. At increasing coverage, dissipation to preadsorbed H atoms favors sticking over recombination[48] in contrast with the dynamics at low coverage for which, due to the large mass-mismatch between a metal and an hydrogen atom, energy dissipation from the H hot-species to surface lattice vibrations is predicted to be slow.[26] In accordance with this, experiments have revealed weak surface temperature effects on H abstraction.[39,42,49] However, to the best of our knowledge, all the above theoretical studies have overlooked any possible effect of electron–hole (e–h) pair excitations on hot-atom abstraction[23,24,28] though their influence has been experimentally suggested.[50] This issue is of prime importance as dissipation to electronic excitations, which is ubiquitous upon interaction/scattering of atoms and molecules with metals,[51–61] was recently predicted to largely dominate the relaxation of light hot-atoms at surfaces,[62–67] which was shown by recent experiments.[68] Consequently, it will significantly affect the hyperthermal motion of hot species.[63,65–67,69] Here we investigate this issue within the framework of hydrogen recombination on W(110).

Among the different theories developed to deal with e–h pair excitations in molecular processes at metal surfaces,[70] the local-density friction approximation[71] (LDFA) offers a good compromise between accuracy and simplicity.[68,72] This approach has already

*a* CNRS, ISM, UMR5255, F-33400 Talence, France.
E-mail: pascal.larregaray@u-bordeaux.fr
*b* Université de Bordeaux, ISM, UMR 5255, F-33400 Talence, France
*c* Donostia International Physics Center (DIPC), Paseo Manuel de Lardizabal 4, 20018 Donostia-San Sebastián, Spain
*d* Instituto de Física Rosario (IFIR) CONICET-UNR, Esmeralda y Ocampo, 2000 Rosario, Argentina
*e* Departamento de Física de Materiales, Facultad de Químicas (UPV/EHU), Apartado 1072, 20080 Donostia-San Sebastián, Spain
*f* Centro de Física de Materiales CFM/MPC (CSIC-UPV/EHU), Paseo Manuel de Lardizabal 5, 20018 Donostia-San Sebastián, Spain31378 | *Phys. Chem. Chem. Phys.*, 2016, **18**, 31378–31383    This journal is © the Owner Societies 2016

served to investigate dissipation in gas-surface elementary processes.[63,65–69,73,74] Making use of this methodology, we here investigate the energy dissipation due to e–h pair excitations in the abstraction of light H atoms on a H-covered W(110) surface. The recombination of hydrogen on tungsten is a system of interest, as this metal is the main candidate for the diverters of the ITER fusion reactor.[11,12] Besides, H-covered W(110) surfaces have generated interest because of the substrate phonon anomalies observed at near saturation coverage[75,76] and the production of highly excited molecules.[14,15,17]

The normal incidence scattering of atomic hydrogen off a H-covered W(110) surface is simulated using quasi-classical trajectories (QCTs) that include the zero point energy (ZPE) of the adsorbates, for $\Theta = 0.25$ ML and 1.0 ML coverages. The approach relies on a DFT based multiadsorbate potential energy surface (PES). It was developed by Pétuya et al.,[48] from an adaptation of the corrugation reducing procedure (CRP)[77–79] describing the ER abstraction dynamics[80] within the frozen surface approximation. Related DFT calculations were carried out using the slab supercell approach and within the generalized gradient approximation proposed by Perdew and Wang (PW91)[81,82] to describe electronic exchange and correlation. The finite coverage potential is developed as a two-H term expansion, i.e.,

$$V(\{\mathbf{r}_i\}) = \sum_{i=1}^{N} V^{3D}(\mathbf{r}_i) + \sum_{i=1}^{N} \sum_{j>1}^{N} I^{6D}(\mathbf{r}_i, \mathbf{r}_j), \quad (1)$$

where $\mathbf{r}_i$ is the position vector of atom $i$, $V^{3D}(\mathbf{r}_i)$ is the three-dimensional H/W(110) interaction potential, and $I^{6D}(\mathbf{r}_i,\mathbf{r}_j)$ is the six-dimensional diatomic interpolation function.[80] The PES developed in ref. 48 has been generalized to describe H penetration to the subsurface down to $-4.4$ Å ($Z = 0$ is defined by the altitude of the topmost surface layer). QCT calculations were performed using a $6 \times 6$ rectangular ($a \times a\sqrt{2}$) supercell with periodic boundary conditions in order to model an infinite covered surface. The classical equations of motion are integrated for one projectile atom and 18 (72) adsorbed targets for $\Theta = 0.25$ ML (1.0 ML) coverages, respectively. The targets initially sit in their equilibrium positions and are given the ZPE as detailed in previous works.[80,83] The initial altitude of the projectile is taken in the asymptotic region of the potential, at $Z_p = 7.0$ Å from the surface. The $(X_p,Y_p)$ initial position of the projectile is randomly sampled in the covered surface irreducible unit cell (yellow areas in Fig. 1). For $\Theta = 0.25$ ML (1.0 ML) coverage, 1 200 000 (300 000) trajectories have been computed to ensure convergence. Since the multiadsorbate PES ignores possible interaction between three hydrogen atoms (see eqn (1)), trajectories are stopped whenever one H atom has two neighbouring H atoms closer than 1.5 Å. As the actual fate of such trajectories is unknown, the corresponding contribution is taken as an uncertainty to any possible outcome of scattering defined below.[48,80,83]

In order to rationalize nonadiabatic effects upon scattering, molecular dynamics simulations are performed within the Born–Oppenheimer Static Surface (BOSS) approximation and within the LDFA. In the former model, neither energy exchange with the surface phonons nor electronic excitations are accounted for. In the latter, electronic nonadiabaticity is introduced through a dissipative force in the classical equations of motion for the hydrogen atoms. Such a frictional force, proportional to the atom velocity, depends on the vector position, $\mathbf{r}_i$, of the atom $i$, through the friction coefficient, $\eta(\mathbf{r}_i)$. Within the LDFA, $\eta(\mathbf{r}_i)$ is taken as that of the same atom moving in a homogeneous free electron gas with electronic density equal to that of the bare surface at the same $\mathbf{r}_i$ position. This approximation has been used to rationalize the relaxation of H hot-atoms on a bare metallic surface,[63,65,67] as well as for recombinative $H_2$ desorption induced by fs-laser pulses.[84] To prevent leakage of the ZPE, the friction acts only when the energy of the preadsorbed atom exceeds the ZPE. Dissipation to surface phonons is here ignored on the ground that, as recently demonstrated,[62–65] dissipation to electrons largely dominates the relaxation of hydrogen on metals at short times ($<500$ fs), which is of interest in the following.

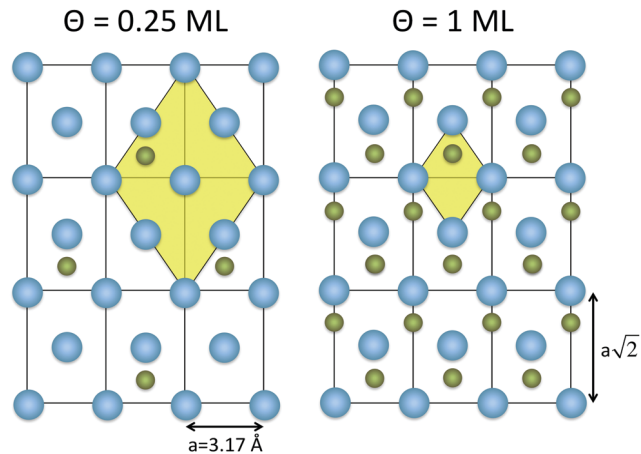

Fig. 1 Positions of the adsorbed H atoms (green points) at $\Theta = 0.25$ ML (left) and $\Theta = 1.0$ ML (right). The lattice constant parameter is $a = 3.17$ Å. Sampling areas of initial $(X_p, Y_p)$ positions of the projectiles for each coverage are represented in yellow.

As a result of H-atom scattering, the following exit channels are defined: reflection whenever one atom reaches the initial altitude of the projectile, abstraction if a $H_2$ molecule reaches this altitude, absorption if one atom lies below the surface ($Z < 0$ Å) after the 1 ps total integration time and otherwise adsorption. It has been checked that a longer integration time does not change the results. The Eley–Rideal (ER) abstraction is assumed to occur when the formed molecule moves definitively toward the vacuum after the first rebound of the projectile.[85] Otherwise, abstraction is considered as the primary hot-atom (HA) process. Following these definitions, the primary HA channel encompasses recombination after hyperthermal travelling of the projectile as well as possible abstraction resulting from the collision of the projectile in the vicinity of the target but leading to several rebounds of the center-of-mass of the forming $H_2$ molecule. When abstraction takes place involving two target atoms, it is classified as a secondary HA process.[32]

The cross-sections per adsorbate for the above-defined exit channels are displayed in Fig. 2 and 3 as a function of the incidence energy of the projectile, $E_{in}$, at $\Theta$ = 0.25 ML and $\Theta$ = 1 ML coverages, respectively. For both coverages, the effect of e–h pair excitations is to increase adsorption at the expense of absorption, reflection and abstraction. Nevertheless, the qualitative evolution of these channels with $E_{in}$ hardly changes. At low $E_{in}$, most of the projectiles adsorb on the surface whereas absorption and reflection significantly increase with $E_{in}$.[48] The reduction experienced by the abstraction cross-section when accounting for e–h pair excitations is much more pronounced for low coverage and low $E_{in}$ and, as apparent from the lower panels of Fig. 2, mainly originates from a drastic reduction in the HA channels. At 1 ML coverage, the effect of e–h pair excitations on abstraction is smaller (see Fig. 3). Besides, as previously highlighted,[85] electronic excitations slightly affect ER abstraction, a result that has been recently rationalized in terms of effective reduction of $E_{in}$.[85]

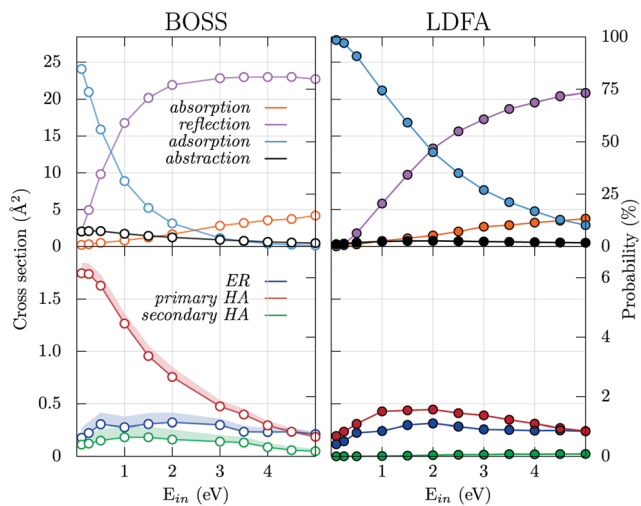

Fig. 2 Top panels: Cross-sections per adsorbate for adsorption (blue), absorption (orange), reflection (purple) and abstraction (black) as a function of the projectile incidence energy, $E_{in}$. Bottom panels: Cross-section per adsorbate for ER (blue), primary HA (red) and secondary HA (green) as a function of $E_{in}$. The numbers at the right axis represent the corresponding probabilities. Left (right) panels correspond to BOSS (LDFA) results. The surface coverage is 0.25 ML. Uncertainties, which correspond to the contribution of stopped trajectories (see the text), are represented by shaded domains when their contribution is larger than the size of the symbols.

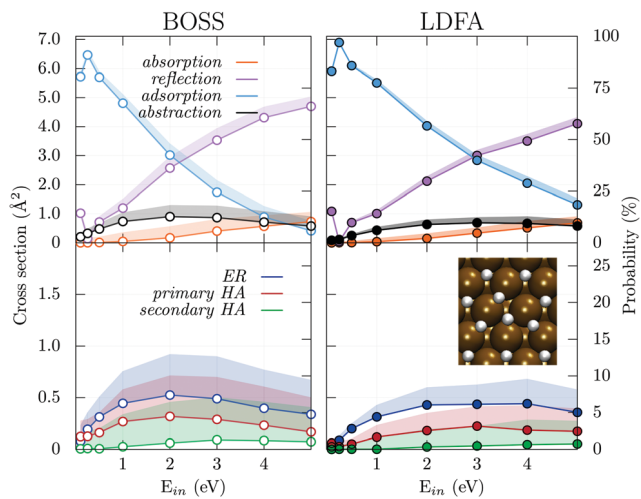

Fig. 3 Same as in Fig. 2 but for 1 ML surface coverage. Inset: Adsorption structure from DFT (see the text).

The already small contribution to abstraction *via* secondary HA decreases when including electronic excitations, particularly at $\Theta$ = 0.25 ML, for which this contribution almost disappears. Overall, within the LDFA approximation ER and HA mechanisms compete whatever the coverages, but as the coverage increases ER becomes the dominant abstraction channel.

To understand the significant decrease of the primary HA process at $\Theta$ = 0.25 ML and $E_{in}$ = 0.5 eV, we have followed the time evolution of the total energy of the projectile, $E_p$, for both BOSS and LDFA simulations. At each time step, $E_p$ is calculated at the projectile position $\mathbf{r}_p$ as

$$E_p = K_p + V^{3D}(\mathbf{r}_p) + \frac{1}{2}\sum_{i \neq p}^{N} I^{6D}(\mathbf{r}_i, \mathbf{r}_p), \quad (2)$$

where $K_p$ is the kinetic energy of the projectile and its potential energy is approximated as the sum of the surface-projectile atom potential $V^{3D}(\mathbf{r}_p)$ and half of the diatom interpolation potential $I^{6D}(\mathbf{r}_i,\mathbf{r}_p)$, which describes the projectile–adsorbate interaction on the surface. Fig. 4 (right panel) displays the calculated $E_p$ distribution and the fraction of the projectiles still traveling on or below the surface ($-4.4$ Å $< Z_p < 3.5$ Å) at increasing integration times. The results are obtained in each simulation from 100 000 trajectories. In order to facilitate the implications of $E_p$ in determining whether the projectile will permanently be trapped on the surface, the left panel sketches the potential energy diagram for the adsorption and abstraction processes. The origin of potential energy is chosen for the projectile at infinite distance from the covered surface and the adsorbates sitting at their equilibrium position. Both adsorption and abstraction processes are exothermic by about 3 and 1.5 eV, respectively. The red line represents the initial total energy of the projectile with $E_{in}$ = 0.5 eV incidence energy ($E_p = E_{in}$). The total energy must hence decrease from 0.5 down to −3 eV for the projectile to stick in a surface three-fold site. However, abstraction already becomes endothermic as soon as the projectile energy $E_p$ is lower than −1.5 eV. Comparing the BOSS and LDFA energy loss distributions plotted in the right panel, we find that when e–h pair excitations are included in the calculation (black), 80% of the projectiles already have an energy below −1.5 eV after 0.2 ps. Consequently, these atoms cannot lead to recombination and adsorbed. In contrast, the BOSS energy distribution shows that after 0.2 ps more than half of the projectiles still have enough energy to recombine with an adsorbate. In this case, 0.8 ps is required in order for all projectiles to lose 2.0 eV. As a consequence, at $\Theta$ = 0.25 ML coverage, the energy loss due to e–h pair excitations highly reduces the recombining hot-species lifetime. This is illustrated in Fig. 5, where the distributions of ER and primary HA abstraction times as obtained within the BOSS and LDFA simulations are displayed. This time is taken as the total time for abstraction. When accounting for e–h pair excitations, the timescale for both abstraction processes become really similar and, concomitantly, the distances travelled on the surface before recombination. Actually, the abstracted adsorbates are basically the ones initially located in the irreducible surface unit cell or in the first periodic cells (not shown). The reduction of the projectile traveled length caused by e–h pair excitations is also the reason of the strong decrease observed in the LDFA secondary HA probabilities, in particular at low coverage. As one can notice from Fig. 3, the influence of e–h pair excitations for high coverages ($\Theta$ = 1 ML) is less important. On the one hand, previous BOSS simulations performed at such coverages[48] predicted that the projectile energy is efficiently dissipated into the other adsorbates, resulting in a short lifetime for the hot species. On the other hand, a high density of adsorbates also results in a screening of the impinging hydrogen which then cannot interact closely with the surface. The energy loss mediated by e–h pair excitations thus influences much less reactivity at such coverages.

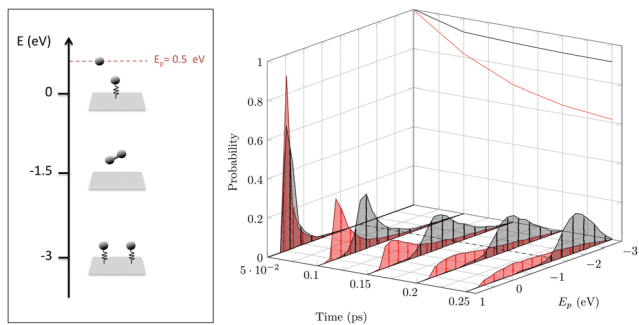

Fig. 4 Left panel: Potential energy diagram for adsorption and abstraction processes. Energies are in eV. Right panel: Total energy distributions of the projectiles travelling at the surface at different times for BOSS (red) and LDFA (black) calculations for $E_{in}$ = 0.5 eV and $\Theta$ = 0.25 ML (left front plane, arbitrary units). The curves in the right front plane display the fraction of projectiles located at heights $-4.4$ Å $< Z < 3.5$ Å as a function of time. The dashed line ($E_p = -1.5$ eV) indicates the threshold energy below which abstraction becomes endothermic.

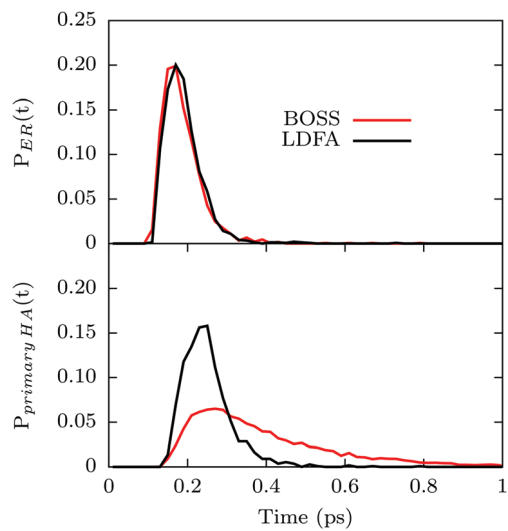

Fig. 5 Normalized distribution of ER (upper panel) and primary HA abstraction times within the BOSS (red) and LDFA (black) simulations at 0.25 ML coverage.

At high coverage ($\Theta$ = 1 ML) and low incidence energy (<0.5 eV), the simulations predict a large adsorption probability (>80%), suggesting possible supersaturation of the surface. The dynamics predict that the extra atom adsorbs in the second three-fold hollow site of the unit cell. In order to discard any interpolation error, additional DFT relaxation calculations have been carried out for 41 adsorbates ($\Theta$ = 1.025 ML) using a 5 × 4 surface unit cell and the same parameters previously adopted for the construction of the CRP PES.[86] These calculations confirm indeed that the adsorption of an H atom on the saturated surface is energetically favourable by 2.07 eV, the structure of the surrounding H atoms being slightly distorted as illustrated in the inset of Fig. 3 (right). Interestingly, such adsorbing sites with lower binding energy have already been anticipated in the literature to rationalize the experimentally observed hot vibrational state distributions of $H_2$ molecules resulting from abstraction on tungsten surfaces.[16,17] Low-energy electron diffraction and inelastic He-atom scattering measurements have actually confirmed the formation of a hydrogen superstructure at $\Theta$ = 1.5 ML coverage.[87]

To conclude, we have theoretically investigated for the first time the effect of e–h pair excitations on hydrogen abstraction on a metal surface at finite coverage. Surprisingly, the HA recombination mechanism, which is supposed to dominate recombination at low coverage and $E_{in}$, is predicted to be significantly affected by the low-energy electronic excitations, since they greatly reduce the relaxation time of hot hydrogen on the W(110) surface. As a result, the HA mechanism is considerably diminished in favor of H adsorption for these incidence conditions. This work thus sheds new light on the competition between ER and HA abstraction mechanisms of hydrogen on metals.

## Acknowledgements

O. G., J. I. J, and M. A. acknowledge financial support by the Basque Departamento de Educación, Universidades e Investigación, the University of the Basque Country UPV/EHU (Grant No IT-756-13), and the Spanish Ministerio de Economía y Competitividad (Grant No. FIS2013-48286-C2-2-P). O. G., M. A., and P. L. acknowledge the IDEX Bordeaux (ANR-10-IDEX-03-02) and Euskampus for funding. Computational resources were provided by the DIPC computing center, the Mésocentre de Calcul Intensif Aquitain (MCIA). France Grilles is also acknowledged for providing computing resources on the French National Grid Research. This work was conducted in the scope of the Transnational Common Laboratory "QuantumChemPhys: Theoretical Chemistry and Physics at the Quantum Scale".